\documentclass{webofc}
\usepackage[varg]{txfonts}   % Web of Conferences font
\usepackage{subcaption}
\usepackage{romannum}
\begin{document}

%\title{Computational workflow of the LZ dark matter detection experiment}
\title{Optimization of Software on High Performance Computing Platforms for the LUX-ZEPLIN Dark Matter Experiment}

\author{\firstname{Venkitesh} \lastname{Ayyar}\inst{1}\fnsep\thanks{\email{vpa@lbl.gov}} \and
	\firstname{Wahid} \lastname{Bhimji}\inst{1} \and
	\firstname{Maria Elena} \lastname{Monzani}\inst{2} \and     % etc.
	\firstname{Andrew} \lastname{Naylor}\inst{3} \and
        \firstname{Simon} \lastname{Patton}\inst{1} \and
	\firstname{Craig E. } \lastname{Tull}\inst{1} 
}

\institute{Lawrence Berkeley National Laboratory, 1 Cyclotron Rd, Berkeley, California 94720, USA. 
\and
           SLAC National Accelerator Laboratory, 2575 Sand Hill Rd, Menlo Park, California 94025, USA.
\and
           University of Sheffield, Sheffield S10 2TG, UK.}

\abstract{%
 High Energy Physics experiments like the LUX-ZEPLIN dark matter experiment face unique challenges when running their computation on High Performance Computing resources. In this paper, we describe some strategies to optimize memory usage of simulation codes with the help of profiling tools. We employed this approach and achieved memory reduction of 10-30\%. While this has been performed in the context of the LZ experiment, it has wider applicability to other HEP experimental codes that face these challenges on modern computer architectures.}
\maketitle
\section{Introduction} \label{intro}
The LUX-ZEPLIN (LZ) experiment is a direct detection search for cosmic WIMP dark matter particles~\cite{LZ-1,LZ-2}. It is expected to start taking data in 2020. LZ is located in the Davis Cavern at the 4850-foot level of the Sanford Underground Research Facility in Lead, South Dakota, USA. The collaboration consists of more than 37 institutes in five countries, having two data centers in the US and UK.
In this paper, we describe mitigation schemes for some of the computational challenges faced by the collaboration.

The Offline software for LZ is based on the widely used High Energy Physics (HEP) frameworks: Geant4 for simulations~\cite{geant} and Gaudi for reconstructions~\cite{gaudi}. The main simulation package is called {\bf BACCARAT}~\cite{baccarat}. Working above the Geant4 engine, it is used to produce accurate simulations of the LZ detector response to dark matter signals and backgrounds. The output is written in ROOT data format~\cite{root}. {\bf DER } is the collaboration's Detector Electronics Response package that reads the output of BACCARAT and simulates the signal processing done by the front-end electronics. {\bf LZap} is the collaboration's data processing and reconstruction software package. Extracting the charge and time information of the photomultiplier tubes, it applies corrections and calibrations to produce the output required for physics analysis. All LZ software is maintained through a GitLab repository~\cite{gitlab}. Software distribution is achieved using the Cern VM File System (CVMFS)~\cite{cvmfs}.

The LZ US data center is located at the National Energy Research Scientific Computing Center (NERSC) in Berkeley Lab.
The US data center recently ported its workflow from High Throughput Computing (HTC) resources to High Performance Computing (HPC) resources like the {\bf Cori} supercomputer~\cite{cori} at NERSC. The simulation codes (BACCARAT) are quite memory and time intensive, and hence there is special emphasis on monitoring and improving their performance and efficiency on Cori.

\section{Memory optimization}
Cori is composed of two systems: the first consists of 2388 Intel Xeon {\it Haswell} processor nodes and the second consists of 9688 Intel Xeon Phi Knight's Landing or {\it KNL} nodes.
Prior to September 2018, the average of the peak of the memory usage of a typical BACCARAT job was assessed to be at least 5GB. Since the total memory per node on the Cori-Haswell machines is 128GB, this restricted the number of parallel jobs on a single node to 25. As each of these nodes have effectively 64 cores per node (due to hyper-threading), more than twice the number of jobs could run be on each node if we could reduce the memory footprint. Similarly for each Cori-KNL node with 96GB RAM and 68 cores, allocating 5GB of memory per job represented utilization of only 19 of the effectively available 272 threads, since each core uses four threads.
To increase utilization of cores on these machines, this paper reports our approach to optimize the memory usage.

We used two profiling tools to assess memory usage and help identify memory leaks. These are: 
\begin{enumerate}
\item {\bf PRocess MONitor (prmon):}
{\it Prmon}\footnote{https://github.com/HSF/prmon} is a lightweight tool developed by the ATLAS collaboration~\cite{prmon}. It gives a log of memory usage and a few other quantities as a function of time for a given process and all its derived processes. {\it Prmon} is a separate process that runs along with the job.

\item {\bf VTune:}
VTune is a comprehensive profiler developed by Intel~\cite{vtune}. It gives an in-depth profile of the job, detailing the memory usage of individual classes and functions. This package links specific code lines to memory usage, enabling identification of problematic code. Profiling a job with VTune requires the whole node and the run times are considerably longer. %Fig. \ref{Vtune2} gives a sample snapshot of VTune output.
\end{enumerate}

\section{Results}

\begin{figure}[!htb]
\centering
\sidecaption
  \includegraphics[width=.6\textwidth]{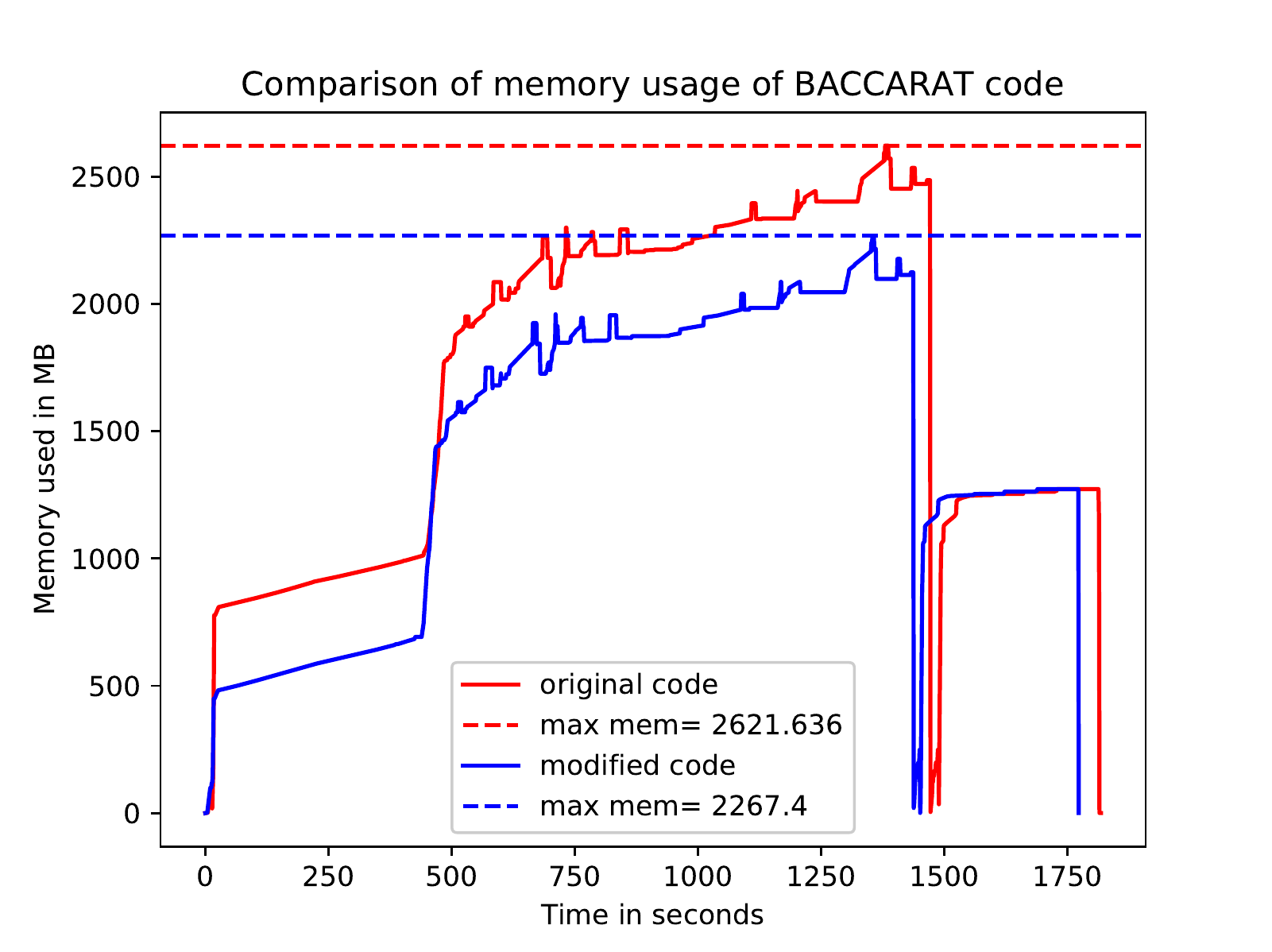}
  \caption{Comparison of memory usage of BACCARAT code before and after modifying the code to fix a memory leak. The X axis shows the wall time while the Y axis denotes the memory usage. The code has three main phases: {\it creation of the event list}, {\it event generation} and {\it clustering and output}. In this case, the change in code reduced the memory usage in the first part of the code, resulting in memory savings of about 350MB.}
  \label{fig:mem_comparison}
\end{figure}

\begin{figure}[!htb]
\centering
\sidecaption
  \includegraphics[width=.6\textwidth]{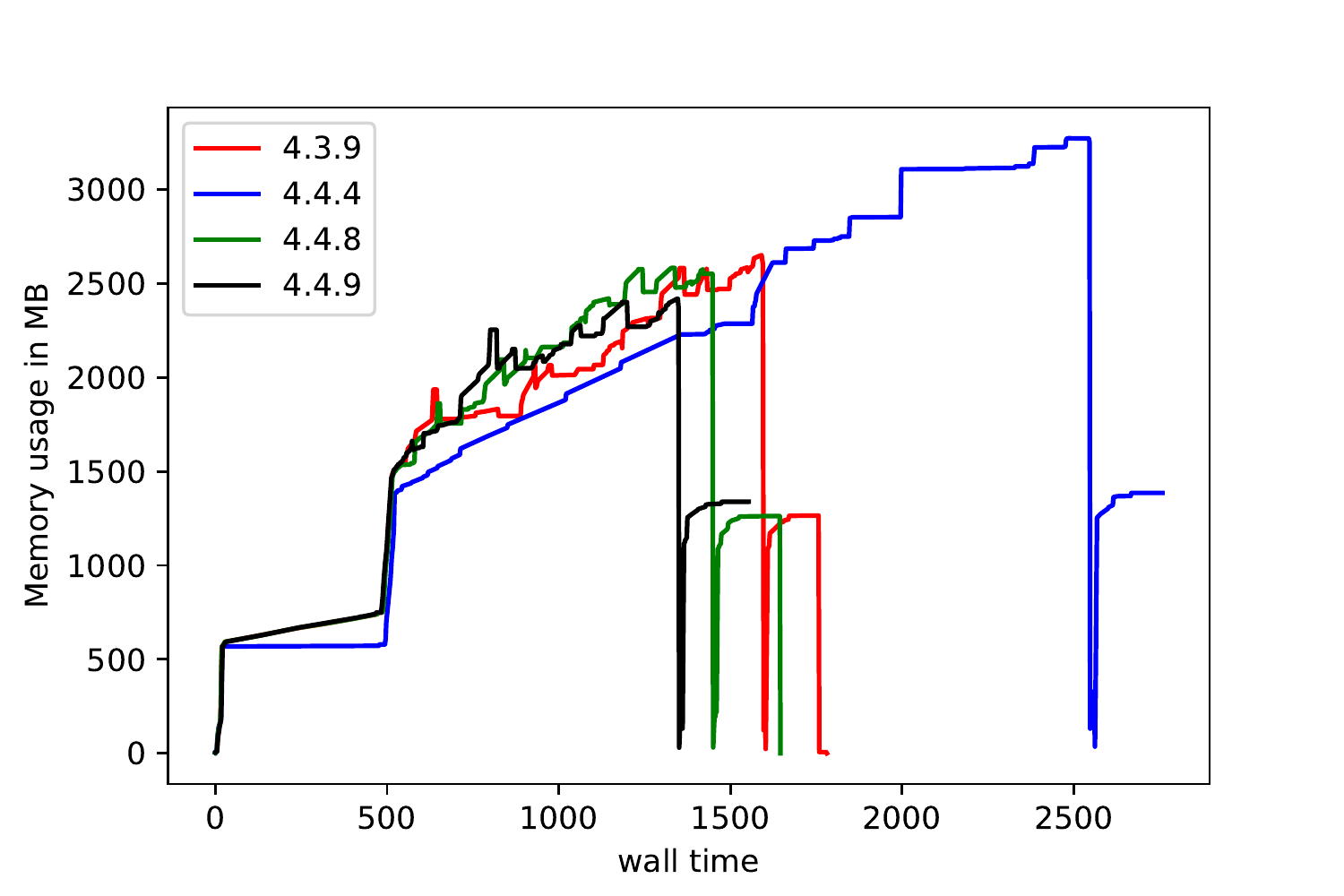}
  \caption{Memory usage in MB for runs with the same initial parameters, for different versions of BACCARAT as a function of wall time. The legend numbers denote the version number of BACCARAT code. The profiling tools helped us confirm the overall behavior of the code in time, thus enabling us to spot anomalous versions such as the blue curve (version 4.4.4) in the above figure.}
  \label{fig:version_comparison}
\end{figure}

\begin{figure}[!htb]
\centering
\sidecaption
  \includegraphics[width=1.0\linewidth]{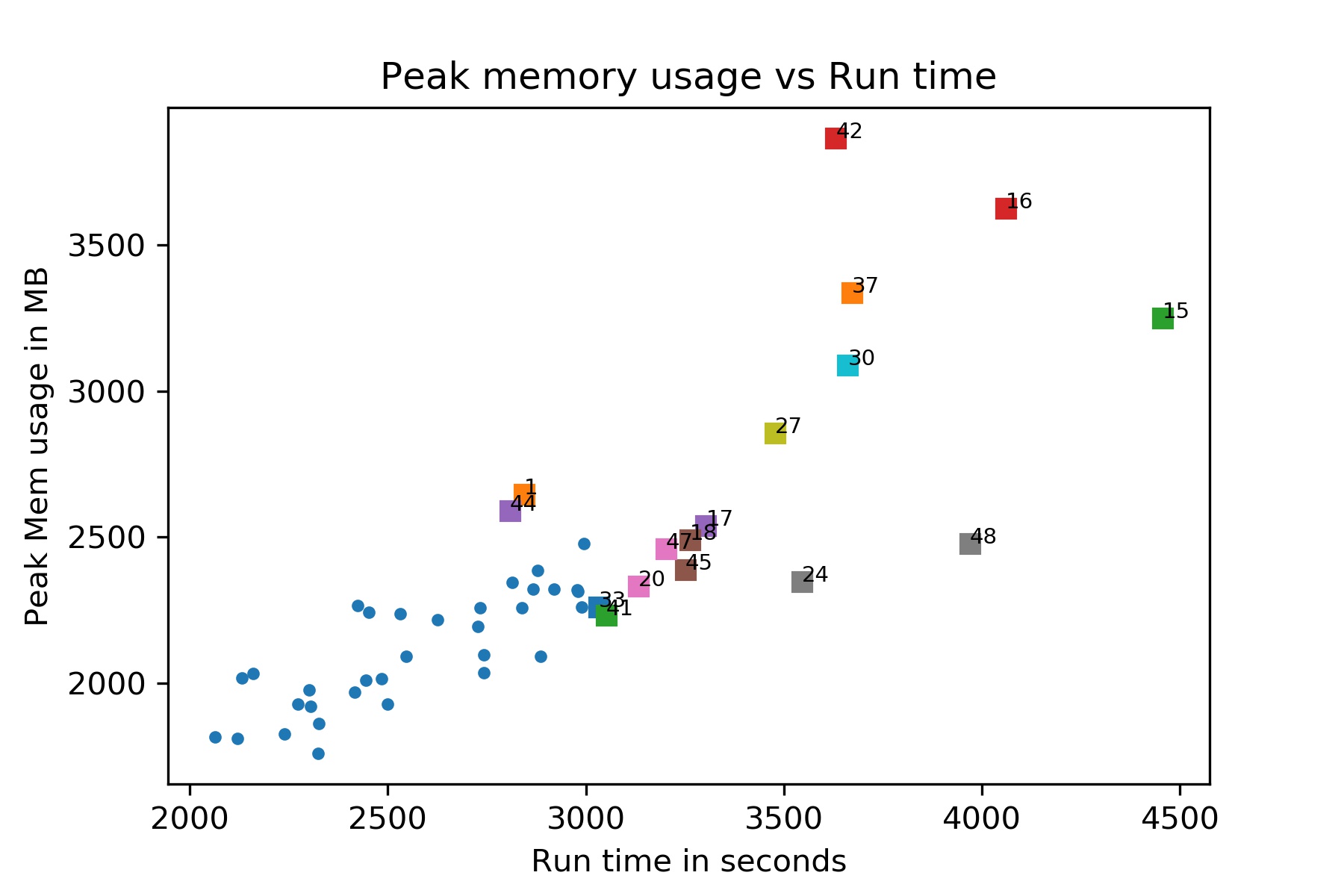}
\caption{This is a scatter plot showing the job run-time on the X-axis and the peak of the memory usage on the Y-axis, for a set of 50 BACCARAT jobs (version 4.9.9) with different seeds, run concurrently on a single node. The circular markers indicate jobs taking less than 2.5GB of memory and having a run-time of less than 3000 seconds. These correspond to about 66\% of the jobs. The square markers correspond to the other 34\% of the jobs and the number next to each symbol indicates the run index of that job. It can be seen that there is a correlation between higher peak memory usage and longer run time.} 
\label{fig:scatter}
\end{figure}

\begin{figure}[!htb]
\centering
\sidecaption
\includegraphics[width=0.6\linewidth]{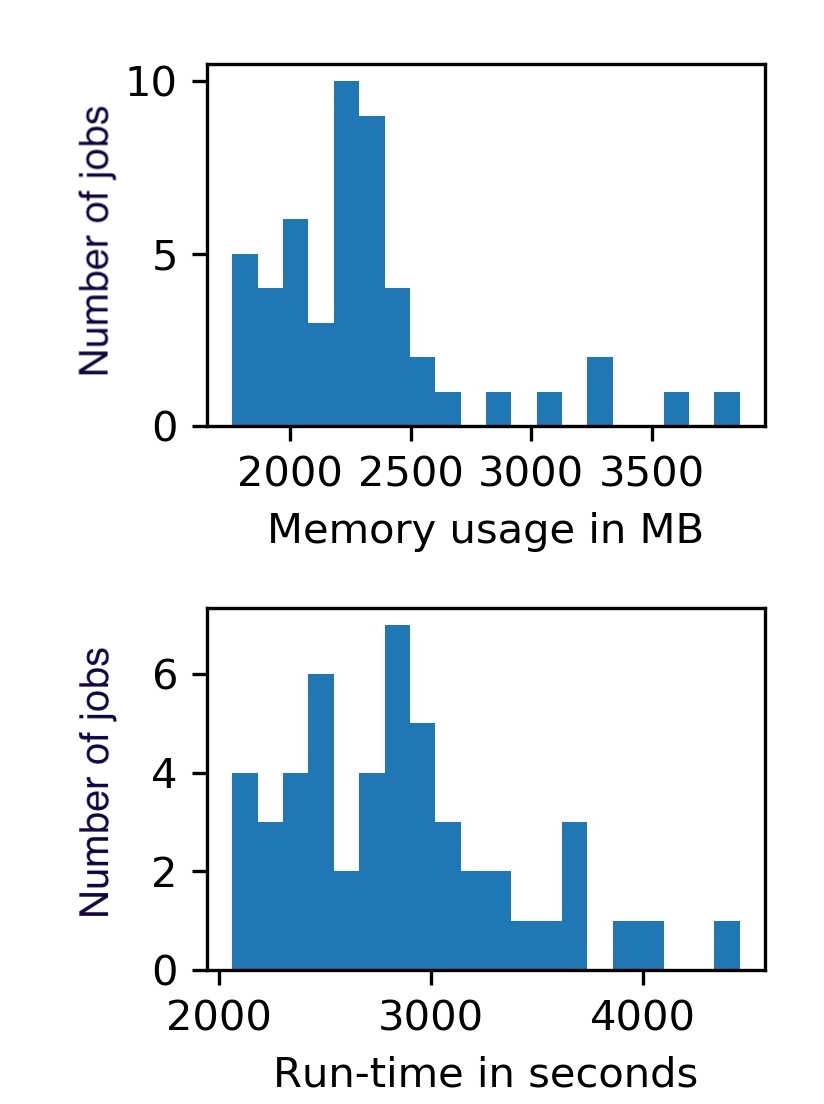}
\caption{ \small Histograms for the peak memory usage and run time for a set of 50 BACCARAT jobs run concurrently on a single node. The top histogram shows the number of jobs on the Y axis and peak memory usage in MB on the X axis. 
The bottom histogram shows the number of jobs on the Y axis and the run-time in seconds on the X axis. 
It can be seen that the average memory required per job is around 2.3GB. }
\label{fig:hist}       % Give a unique label
\end{figure}

Our first step in the memory optimization process is to get an overview of the memory usage of the code using {\it prmon}. Our next step is to get a detailed profile of the memory usage of the code, using VTune. We then modify the code at the locations of possible memory leaks identified by VTune. Having addressed the memory leaks in the code, we 
iterate the process while ensuring the code output remains unchanged. By comparing the memory usage pattern of the two codes, we estimate the improvement in memory allocation and usage. An example is shown in Fig. \ref{fig:mem_comparison}, which compares the memory usage of a BACCARAT version of the code before and after optimization. It can be seen that the change in code reduced memory usage by 350MB.

As the code went through further development, the memory map was obtained using {\it prmon} at each stage. Comparing the memory maps of different versions helped to confirm the overall behavior of the code as it was optimized. This also helped identify code versions with problematic behavior. An example is shown in Fig. \ref{fig:version_comparison}, which compares the memory usage of a few versions of BACCARAT. 

To confirm that this gain in improvement was generally valid, we performed a run of 50 parallel jobs of BACCARAT (version 4.9.9) with different seeds. Each seed represented a different detector condition or event. These were run concurrently on a single node along with {\it prmon} to gather the memory usage for each job. 
Fig. \ref{fig:scatter} is a scatter plot showing the peak memory usage and run time for the 50 jobs. It can be seen that about 66\% jobs are in the lower left corner of the plot corresponding to lower peak memory usage. There is a correlation between higher peak memory usage and longer run time. Fig. \ref{fig:hist} shows histograms of peak memory usage and run time for these jobs. It can be seen that the average memory required per job is around 2.3GB, hence the number of jobs that can be run per node on Cori-Haswell should increase to approximately 50.

\section{Conclusions} \label{summary}
We have discussed the use of profiling tools {\it prmon} and {\it VTune} to profile the LZ simulation code. By implementing a procedure of repeated profiling and code improvements, we have achieved significant memory optimizations. These gains have enabled the running of up to 50 independent jobs per node, thus significantly increasing the efficiency of LZ computation on the Cori supercomputer at NERSC.

\section{Acknowledgements}
We would like to thank Quentin Riffard for his assistance in this work. We would also like to thank Charles Leggett for useful discussions on {\it prmon}. 
This work was supported by the Center for Computational Excellence, a project funded by the Computational HEP program in The Department of Energy's Science Office of High Energy Physics. This work was partially supported by the U.S. Department of Energy (DOE) Office of Science and by the U.K. Science \& Technology Facilities Council. Furthermore, this research used resources of the National Energy Research Scientific Computing Center, a DOE Office of Science User Facility supported by the Office of Science of the U.S. Department of Energy.


\begin{thebibliography}{}
%
% and use \bibitem to create references.
%


% Format for Journal Reference
%Journal Author, Journal \textbf{Volume}, page numbers (year)
% Format for books
%\bibitem{RefB}
%Book Author, \textit{Book title} (Publisher, place, year) page numbers



\bibitem{LZ-1}
D. Akerib et al, \textit{Nuclear Instruments and Methods in Physics Research Section A, 163047,(arXiv:1910.09124)}, October 2019. doi: https://doi.org/10.1016/j.nima.2019.163047

\bibitem{LZ-2}
%Mount et al,  \textit{LUX-ZEPLIN (LZ) Technical Design Report , arxiv: 1703.09144}
B. Mount et al, \textit{Technical Design Report, LBNL-1007256. arxiv: 1703.09144}, 2017

\bibitem{geant}
%S. Agostinelli, et al., GEANT4: A Simulation toolkit, Nucl. Instrum. Meth. A506 (2003) 250–303. doi:10.1016/ S0168- 9002(03)01368- 8.
S. Agostinelli, et al., \textit{Nucl. Instrum. Meth. A506 (2003) 250–303}, doi:10.1016/ S0168- 9002(03)01368- 8.

\bibitem{gaudi}
%G. Barrand, et al., GAUDI - A software architecture and framework for building HEP data processing applications, Comput. Phys. Commun. 140 (2001) 45–55. doi:10.1016/ S0010- 4655(01)00254- 5.
G. Barrand, et al., \textit{Comput. Phys. Commun. 140 (2001) 45–55}, doi:10.1016/ S0010- 4655(01)00254- 5.

\bibitem{baccarat}
LUX-ZEPLIN Collaboration, \textit{arxiv:2001.09363}, 2020.

\bibitem{root}
R. Brun, F. Rademakers, \textit{Nucl. Instrum. Meth. A389 (1997) 81-86.}.

\bibitem{gitlab}
Gitlab Inc., https://about.gitlab.com

\bibitem{cvmfs}
%J. Blomer, C. Aguado Sanchez, P. Buncic, A. Harutyunyan, Distributing LHC application software and conditions databases using the CernVM file system, J. Phys.: Conf. Ser. 331 (2011) 042003. doi:10.1088/1742-6596/331/4/042003.
J. Blomer, C. Aguado Sanchez, P. Buncic, A. Harutyunyan, \textit{J. Phys.: Conf. Ser. 331 (2011) 042003}, doi:10.1088/1742-6596/331/4/042003.

\bibitem{cori}
https://docs.nersc.gov/systems/cori/

\bibitem{prmon}
G. Stewart, A.S. Mete, \textit{PrMon}, https://doi.org/10.5281/zenodo.2554202

\bibitem{vtune}
INTEL VTune, https://software.intel.com/en-us/vtune

\end{thebibliography}
\end{document}